\documentclass[12pt,a4paper,twoside,final,notitlepage, leqno]{article}
\usepackage[english]{babel}
\usepackage[T1]{fontenc}
\usepackage{epsfig, graphicx, amssymb}
\usepackage{amsmath,amsthm,epsfig,amsfonts}
\usepackage{float}
\usepackage{color}
\def\br {\break}

\def \smb {{\scriptstyle \bullet }}

\newcommand{\moneq}{\vspace*{-7pt} \begin{equation} \displaystyle }
\newcommand{\moneqstar}{\vspace*{-6pt} \begin{equation*} \displaystyle }
\newcommand{\monendstar}{\vspace*{-6pt} \end{equation*}   }
\newcommand{\monend}{\vspace*{-7pt} \end{equation}   }
\newcommand{\moneqarraystar}{ \begin{eqnarray*} \displaystyle }
\newcommand{\monendarraystar}{ \end{eqnarray*}   }

\definecolor{vertfonce}{rgb}{0.0, 0.5, 0.0}

\def\section*#1{}

\usepackage{fancyhdr}
\renewcommand{\headrulewidth}{0pt}

\begin{document}

\fancypagestyle{plain}{ \fancyfoot{} \renewcommand{\footrulewidth}{0pt}}
\fancypagestyle{plain}{ \fancyhead{} \renewcommand{\headrulewidth}{0pt}}

~

  \vskip 2.1 cm

\centerline {\bf \LARGE  Interception of a bolide ?}

 \bigskip  \bigskip \bigskip

\centerline { \large    Fran\c{c}ois Dubois$^{ab}$}

\smallskip  \bigskip

\centerline { \it  \small
  $^a$  Conservatoire National des Arts et M\'etiers,  Paris, France.}

\centerline { \it  \small
$^b$    Aerospatiale Espace \& Defense, Les Mureaux, France.}

\bigskip

\centerline {09 October 1997
  {\footnote {\rm  \small $\,$ This contribution has been
presented at the 48th International Astronautical Congress in Turin (Italy,
06-10 October~1997) the 09 OCtober 1997
as   International Astronautical Federation (IAF) paper n$^{\rm o}$ IAA.97-IAA.6.4.09.
Edition 17 November 2023.}}}

\bigskip  \bigskip
\smallskip \hangindent=72mm \hangafter=0
{\it  Abraracourcix ne craint qu'une chose :
c'est que le ciel lui tombe sur la t\^ete, mais comme il le dit lui-m\^eme, ``c'est pas
demain la veille~!''}

\hfill   Ren\'e Goscinny [GU61]

\bigskip 
\smallskip \hangindent=72mm \hangafter=0
{\it  In Mayan language, Chicxulub means ``devil's horns''.} 

\hfill  Charles Frankel [Fra96]

\bigskip  \bigskip  
\noindent {\bf \large Abstract}

\noindent
We present  a review of the literature on the subject of a possible collision between the Earth and a meteorite or comet.
We emphasize the global effects when sufficient energy is involved.
We propose several types of human actions adapted to the physico-chemical nature of the collider bolide.

\bigskip 
\noindent {\bf \large R\'esum\'e}

\noindent
Nous proposons une revue bibliographique sur la question  d'une possible collision entre la Terre
et un m\'et\'eorite ou une com\`ete.
Nous  insistons sur les effets  globaux lorsqu'une \'energie suffisante est mise en jeu
et nous proposons plusieurs types d'actions humaines adapt\'ees \`a la nature physico-chimique du bolide collisionneur.




\newpage 
\bigskip    \noindent {\bf \large   1)	\quad Introduction}

\smallskip 
Our question is prompted by three events.  First, the majority of the scientific
community now agrees that the extinction of the dinosaurs and many other living
species 65~million years ago was caused by an asteroid or comet, ten kilometers in
diameter, which collided with the Earth [Koe96].  Secondly, at the beginning of this
century, a 50-meter object hit the Earth in Tunguska, Siberia, with dramatic effects
on the local vegetation [CTZ93].  A third such event occurred on Jupiter three years
ago, when the two-kilometer Shoemaker-Levy comet crashed into the Jovian atmosphere
[Duf94]. 

\bigskip 
Three questions naturally arise: (i) What probability is there for such a catastrophe
to occur on Earth in the future?  (ii) What is the level of a disaster that is
``acceptable'' for mankind?  (iii) What can be done to prevent such an event?

\bigskip 
We stress that a global methodology is required to consider such a problem and refer
the reader to previous work done at Aerospatiale [Dar93] to prevent damage to a space
plane by micrometeroids.  First, a good knowledge of the threat is required. Second,
the effects of collision of an asteroid or comet with the Earth on biological and
economic equilibrium must be considered. Third, the ideal type of space protection
suitable for preventing such a catastrophe must be calculated.

\bigskip    \noindent {\bf \large   2)	\quad Space threat}

\fancyhead[EC]{\sc{Fran\c{c}ois Dubois}}
\fancyhead[OC]{\sc{Interception of a bolide ?}}
\fancyfoot[C]{\oldstylenums{\thepage}}

\smallskip 
It was recently established that the collision of a
meteorite with the Earth could have major global effects for living species on our
planet.  We need only to consider the sudden disappearance of dinosaurs 65~million
years ago.  In 1980, L. and W. Alvarez, F.~Asaro and H.~Michel [Alv80] suggested that
a correlation between this event and the unusual concentration of iridium in the
Cretaceous-Tertiary K/T geological stratum could correspond to the impact of an
asteroid on Earth.  The asteroid was estimated to have a diameter of 10 kilometers
(the same order of thickness as the troposphere).  Assuming a density of 3~metric tons
per cubic meter and a relative velocity of 30 kilometers per second on impact, the
kinetic energy transmitted to the atmosphere by such a meteorite is around
$10^{24}$~joules. 

\bigskip 
Such an asteroid would produce a crater whose diameter, by initial estimate, would be
around 20 times that of the object.  And a crater with a diameter of 150-250
kilometers was effectively (re)-discovered in 1991 by A.~Hildebrand [HBC91] at
Chicxulub in Yucatan, Mexico, with difficulty due to a layer of post-impact sediment
around, 1 km thick.  The correlation between this major biological extinction and the
Chicxulub crater now seems well established. See, for example, the review article of
C.~Chapman and D.~Morisson [CM94], the synthesis made by C.~Frankel [Fra96] and the
scientific proposal of C.~Koeberl {\it et al}. [Koe96].

\bigskip 
Moreover, geological observations suggest that four other similar catastrophic events
occurred during the last 500 million years [CM94]. 

\bigskip 
The 1908 event in Tunguska, Siberia, is more mysterious.  An area of 2000 square
kilometers of forest was destroyed by a meteorite that exploded in the atmosphere
without causing any human victims.  The atmosphere was very perturbed, ``night lights''
were observed in Europe and Asia [Whi30] and, according to C. Sagan [Sag80], the
resulting shock wave traveled twice around the Earth.

\bigskip 
Scientific studies of this event did not begin until 20 years later with the mission
of the geologist L.~Kulik [Kul27].  The energy transmitted to the atmosphere was
estimated at $5 \, 10^{16}$ joules.  But the origin of this meteorite does not seem
to have been fully established.  According to Sagan [Sag80], the Earth's orbit
intersected the orbit of Enke's comet on June 30, 1908, and the meteorite could be a
piece of this comet.  It should be noted that a kinetic energy of
 $5 \, 10^{16}$ joules is compatible with a 100-meter object with a density of 0.1
metric ton/m$^{3}$ (typical for comets [Wyc87]) and a relative velocity of 30 km
per second.  The models proposed by C.~Chyba, P.~Thomas and K.~Zahne [CTZ93] exclude
dense asteroids and light comets but show that the Tunguska meteorite could be a stony
asteroid.  The energy data are consistent with a 50-meter object with a specific
gravity of 4 relative to liquid water and an incoming velocity of 15 km/s.  It is
interesting to note that this relatively ``small'' impacting energy is equivalent to
that obtained with 10 megatons of TNT, {\it i.e.} 1000 times more than that of the atomic
bombs which destroyed Hiroshima and Nagasaki in August~1945. 

\bigskip 	
Here we wish to focus on very recent ``non-events'' concerning interactions between
Earth and small asteroids. The asteroid Toutatis (1989AC) discovered by a French team
of astronomers (A.~Maury {\it et al}. [Mau89]) is a dual object composed of two stones, 4
and 2 kilometers in diameter, that approaches the Earth every four years.  The minimum
distance between Earth and Toutatis was equal to 3.6 million kilometers on December 8,
1992 and will be only 1.5 million kilometers on September 29, 2004 (see {\it e.g.}
[Aug92]). 

\bigskip 
Recently, a small 9-meter object (1991 BA) intersected the Earth's orbit at only
170,000 kilometers (half the distance between Earth and Moon!) [SRM91].  Object 1989FC
with a diameter of 300 meters, which crossed the Earth's orbit at only 700,000
kilometers on March~23, 1989, was more of a hazard (see {\it e.g.} [Fra96]). 

\bigskip 	
Currently, no systematic effort is being made to catalog such small asteroids with
diameters of, for instance, 100 meters to 10 kilometers, orbiting the Sun and in a
position to intersect the Earth's orbit, causing a major catastrophe.  Certain
families of asteroids are known ({\it e.g.} Apollo, Amor, Atens) and some 250 objects have
been identified [Fra96].  But there are an estimated 3000 asteroids in the 1-10 km
class [SWS90]. 

\bigskip 
Detection of small asteroids is not a trivial problem.  For instance, to be compatible
with the diffraction of an electromagnetic wave with a wavelength
$\, \lambda  = 0.5 \,  10^{-6} $ meters, the telescope required to detect a meteoroid with a
diameter $\, \alpha \,$ of 1 km located at a distance $\, L \, $  of 1 astronomical
unit ($150 \,  10^{6} $  kilometers) would have to have a diameter $\, D \,$ 
calculated by the following equation: 

\smallskip
\centerline {$ \displaystyle 
{\rm  tan} \, \theta \,   = \, {{\alpha}\over{L}} \,\, \leq \,\,   1.22 \, {{\lambda}
\over{D}} \,$}

\smallskip \noindent 
which means that the diameter would have to be more than 100
meters!  This unrealistic instrument could be replaced by a sequence of
interferometers using CCD cameras, but this technology is still in development, {\it e.g.}
at Kitt Peak Observatory (see [Spa94]).

\bigskip 
It should also be noted that long-range prediction of the orbits of such small objects
is not obvious because of their complex interactions with the Sun and Earth.  The
trajectory of an object such as Toutatis exhibits a strong resonance correlation with
the Earth and this regular transfer of energy could be potentially hazardous in the
future.  We feel that a special effort should be made to obtain accurate ephemeris
data on these very small objects.

\bigskip 
A general effort should also be made to develop a system for studying the Earth's
environment to be able to determine the real risk of collision of an asteroid with
Earth in the future.

\bigskip 
Comets are a second family of meteorites that can collide with the Earth.  The risk
appears small for periodic comets ({\it e.g.} Halley's comet) but the most interesting
comets are irregular.  The problem is to determine the probability of collision of
such an irregular comet with the Earth over a typical period of  $10^6$ years. 

\bigskip 
We can imagine a stochastic model based on the hypotheses of Oort's cloud (see
[Oor63]) located 50,000 astronomical units = $ 7.5 \,  10^{12}\, $  kilometers from
the Sun.  A comet with mass $\, M \,$  (what is the mathematical probability of this
random variable?) has an initial position (what is the probability of this random
variable?) with a velocity $\, V \,$  (same question) and enters the solar system (what
is the probability for the date of injection?).  When the positions of the planets are
known at time zero (what is the law of this random vector?), it is sufficient (but is
it possible?) to compute the irregular comet's trajectory to determine the probability
of collision with the Earth and the possible impact parameters (for instance, a
relative velocity of such a comet is conventionally estimated at 50-60 km per second). 

\bigskip 
To our knowledge, this modeling problem has not yet been solved.  A major difficulty
probably resides in determining the appropriate approximate model that can take into
account the mean values of the main planets in the solar system and the filtering
effect of Jupiter and Saturn on telluric planets.  Such a theoretical development
combined with knowledge of comets with an irregular or long period for calibration of
the parameters appears important for determining this risk based on irregular comets. 

\bigskip 	
The risk of a collision between the Earth and a meteorite in the future is related to
two separate threats: near-Earth asteroids and long-period comets.  The first threat
is not known, because no systematic program of observations has been developed to
detail all the 100 m-10 km objects approaching the Earth.  The second threat is
structurally uncertain.  In both cases, our knowledge could be improved by modeling to
accurately predict the future possible collision of objects with complicated
trajectories such as the asteroid Toutatis or determine the probability of collision
with an irregular comet.

\bigskip  \newpage   \noindent {\bf \large   3)	\quad Global effects of a single collision}

\smallskip 	
We begin again with the remark that our knowledge of previous global catastrophes
remains limited.  The Cretaceous/Tertiary transition 65 million years ago has been
correlated with a meteorite's impact.  But the exact cause of the disappearance of 70
percent of living species on Earth is still an open question.  Recent studies (R.
Schultz and S. d'Hondt [SH96], N. Bhandari, V. Courtillot and R. Rocchia [BCR97]) show
that coupling between asteroid impacts and volcanic eruptions is considerable.  In
particular, the amount of collision energy transferred to the atmosphere and biosphere
is a very crucial parameter. 

\bigskip 
The energy levels of such collisions must be studied. A Chicxulub-like impact is
associated with the transfer of $10^{24}$  joules from the meteorite to the Earth,
whereas the 1980 eruption of Mount Saint Helen's volcano caused an input of (only!)
$10^{16}$  joules, very comparable to the (small?) Tungunska event of 1908.  It
should also be noted that the total energy released annually by the Earth (heat flow,
volcanoes, earthquakes, etc.) is estimated at\br
$10^{21}$  joules [Koe96]. 

\bigskip 
If $\, 10^{24} $  joules were transferred to the Earth's system, the amount of energy
to be dissipated would amount to $10^{14}$ joules {\it i.e.} the energy of $10^{6}$
100-watt light bulbs per person during one year considering a world population of
$10^{10}$  people. 

\bigskip 
An event comparable to the Chicxulub impact occurred in July 1994, when the
Shoemaker-Levy comet entered Jupiter's atmosphere.  This comet was discovered in 1993
[SSL93] and probably broke up into 20 pieces one year before discovery, when this
long-period comet first entered Jupiter's gravitational field.  The impact of several
objects was predicted theoretically [Sho94] and the effects caused a spot 1900
kilometers in diameter comparable to a giant red spot with light emission due to
excitation of very hot gases.  The total energy transferred by the nine impacts
between July 16 and 22, 1994 is estimated at $ 4 \,\, 10^{21}$  joules and the total
diameter of the initial object was around 2 km. 

\bigskip
This interaction between the incoming meteorite and the Jovian atmosphere is now being
modeled (see {\it e.g.} [Tak95]).  Its effects on Jupiter's meteorology are still large
several years after the collision.  It should be noted that another collision of this
type between a meteor and Jupiter was probably observed by J.D. Cassini in
December 1690 (see {\it e.g.} [Hir97]). 

\bigskip 	
Is it possible to mathematically describe the effects of rapid interactions such as
the above?  We assume that a mathematical model of the atmosphere and biosphere can be
obtained by a dynamical system using differential equations, {\it i.e.} the search for a
vector $\, X(t) \,$ with a finite number of real components (or parameters) as a function
of time, satisfying a nonlinear equation of the type: 

\smallskip 
\centerline {$ \displaystyle 
{{\rm d}\over{{\rm d}t}} \, \bigl(  X(t) \bigr) \,  = \,  f\bigl(  X(t) \bigr) \,  +
\,  g(t) $}

\smallskip \noindent
where $\, f(\smb)\,$  describes all the physical-chemical-biological models
coupling the internal variables of system  $\, X \,$  and  $\, g(\smb)\,$ 
describes the external force, typically a function containing at least two periods, the
24-hour ``day forcing'' and the 365-day ``year forcing''.  An event such as Chicxulub
corresponds to adding an enormous Dirac delta function to term  $\, g(\smb)\,$  in the
right-hand side of the above model.  After integration over a short period, the initial
conditions are suddenly changed at $\, t=0 \,$ when the catastrophic event occurs.  It
seems very reasonable mathematically that different equilibria can be found by the system
as the time approaches infinity after such a large perturbation (see {\it e.g.}  [HS74] 
for the mathematical foundations of this type of dynamical phenomenology).  In sum, the
impact effects are hazardous if they create ruptures in the global equilibria of the
Earth's ecological system. 

\bigskip 	
The community of Earth sciences has developped a Global Change program (see {\it e.g.}
[GC96]) whose objective is the development of a good understanding of global
modifications on biosphere, atmospheric chemistry, terrestrial ecosystems, etc. that
are induced by human activities on the planet. This research program is
extraordinarily complex and ambitious but mathematically speaking, it corresponds to
regular modifications of function  $\, g(\smb) $.   On the contrary, the project of
new program of Response of the Earth System to Impact Cratering [Koe96] is much more
adapted to these abrupt modifications. We think that this programme is scientifically
essential and could be incorporated in a much more complex one. In fact, actual
studies concerning risks induced by impact on the Earth by asteroids are at our
knowledge based on a life-death binary option [CM94]. Our intuition is that reality of
a global catastrophe is much more complex to describe. The important fact is to
consider the relations between human beings and a global catastrophe can occur whereas
few persons are killed by the initial cause like the impact of an asteroid but if the
induced effects destroy the essential links that makes economical activity
(agriculture, industry, treatment of information).

\bigskip 
Here we wish to focus on the necessity of studying the global effects of collision of
a meteorite with the Earth using coupled methods that take into account, not only
physically, spectacular phenomena such as those observed with the Shoemaker-Levy
comet's impact on Jupiter, but also the biological and economic consequences.  In
particular, to be ``acceptable'', a collision must not damage either the essential links
between human beings (energy fluxes, hydrology) or the more complex ones related to
communication, such as education, economy and state organization.  The availability of
detailed coupled knowledge could be used to determine the cost of an intervention
system.

\bigskip    \noindent {\bf \large   4)	\quad Ideas for Human Action}

\smallskip 	
First, it should be remembered that the collision energy of a meteorite with the Earth
is basically kinetic, with relative velocities of 10 to 70 kilometers per second. 
Below, we consider two typical meteorites, modeling a stony asteroid and a long-period
comet. 

\bigskip 
The first object is a cube with length $\, L = 100 $  meters, density $\, \rho  =
3 $ metric tons/m$^{3}$  and relative velocity  $\, V_1 = 20 $ km/second.  These
parameters induce a mass $ \,  M_1 =  \rho \, L^3 =   3 \, 10^{6}$ metric tons and a
kinetic energy on impact $\,\, W_1   = {{1}\over{2}} M_1 \, V_1 ^2  = 6 \, 10^{17}$
 joules.

\bigskip 
The second object is Halley's comet, for which the data were synthesized by S. Wyckoff
[Wyc87]: a sphere with a half-surface area of 100 km$^2$ ({\it i.e.} a diameter of 8
km) and a density $\, \rho = 0.1 \,$ metric tons/m$^3$ consistent with the icy conglomerate
model for comet nucleus proposed by F. Whipple in 1950 (see [Whi63]).  The mass of
Halley's comet can then be approximated at $\, M_2  = 3 \, 10^{10} $  metric tons,
and the typical relative velocity for long-period comets is around $\, V_2
 = 50 $ km/second [CM94].  The kinetic energy $\, W_2 = {{1}\over{2}}\, M_2 \, V_2^2
\,$   is then equal to $\, 4 \,\, 10^{22} \, $ joules, {\it i.e.}
10 times that of Shoemaker-Levy's comet and only 1/25th that of Chicxulub. 

\bigskip
The first idea that comes to mind is to destroy the asteroid before impact by an
explosion.  But although such an action would transfer some of the energy by changing
the structure of the incoming asteroid, it is not at all clear whether this would
change the total impulse.  For instance, for a 10-metric-ton projectile with a
velocity of 40 km/s, the momentum is  $\, i = 4 \,\, 10^8 $  kg.m/s, which is
negligible compared to the momentum of the small cube
$\, I_1 = M_1 \, V_1 = 6 \,\, 10^{13} \,$   kg.m/s  or the momentum  $\, I_2 = M_2 \, V_2 = 1.5 \,
10^{18} \,$   kg.meters/second  of Halley's comet.  The
center of gravity then continues its initial trajectory and a huge collision is
replaced by a series of smaller ones (see {\it e.g.} J. Ahrens and A. Harris [AH92]).  The
energy transferred to the atmosphere would be considerably decreased by dispersion,
but the remaining collisions are still very hazardous.  According to [AH92],
fragmentation is a satisfactory option only if the maximum fragment size is less than
10 meters so that the fragments burn up in the atmosphere.  In sum, we feel that
fragmentation can only be considered as a ``degraded mode'' of action, when deflection
is not possible. 

\bigskip 	
We now show that deflection of the stony cube is possible using conventional
propulsion technologies.  First, if it occurs sufficiently before the predicted time
of collision, {\it e.g.} one year, a deflection $\, \Delta v = 1 $ cm/s will produce a
much larger displacement due to gravitational amplification ({\it e.g.} by a factor of 20)
than the $\, \Delta x = $ 310 kilometers produced by the simple hypothesis of a straight
trajectory.  Being optimistic about gravitational amplification, we keep this hypothesis
of 1 cm/s as initial velocity deflection. 

\bigskip 
It should be noted here that the kinetic energy associated with this displacement is\br
$ E_1 = {{1}\over{2}} \, M_1 \, \Delta v^2 = 1.5 \,\, 10^5 \, $   joules.  Let us
consider the above small asteroid as a space rocket with a propellant mass $\, m \,$  and
a specific impulse Isp associated with a gas ejection velocity $\, g_0 \, {\rm Isp} = 
 3 \,$  km/s.  The conventional relation (see {\it e.g.} [DF94]): 

\smallskip \centerline {$ \displaystyle 
\Delta v =  g_0 \,\, {\rm Isp} \,\, {\rm log } \Bigl[ {{M_1 + m}\over{M_1}} \Bigr] $}

\smallskip \noindent
simplifies since $\, m \ll M_1 \,$    yielding  $\, m = M_1 \, {{\Delta v }\over{g_0
\, {\rm Isp}}} ,\,$   {\it i.e.} for a cubic asteroid and
the above data, $\, m = $ 10 metric tons of propellant.  It is then necessary to apply such
a push to the asteroid after a complex space rendez-vous somewhere in solar system
relatively close to the Earth's orbit.  This is of course not an easy mission, but it
is a priori compatible with today's technology.  However, detailed design of such a
system remains to be done. 

\bigskip 	
We now consider the problem of giving a similar tap to Halley's comet.  First, it
should be noted that the kinetic energy associated with this deflection of 1 cm/s is
$\, E_1 = {{1}\over{2}} \, M_2 \, \Delta v^2  = 1.5 \, 10^9 \,$   joules [a thousand
megajoules!].  This is an enormous amount of energy from the standpoint of today's
capabilities.  It should be noted that although this energy is huge, it is nothing
compared with $\, W_2 = 4 \,\, 10^{22} \,$   joules that would be transferred to the
Earth's system by a direct collision.  The problem is to create an impulse
$\, J_2 = M_2 \, \Delta v = 3 \, 10^{11} \,$   kg.meters by second somewhere
around Jupiter, because there is a good probability that such a space threat would
interact first with Jupiter, which is a good place in the solar system for playing with
gravity. 

\bigskip 
Aside from the fact that the associated propellant mass ($m = 10^5$  tons) is
unrealistic, the solution suggested above for a small asteroid is obviously unfeasible
because a comet-like object with low density ($\rho = 0.1$) is somewhat like cotton
wool and is a very fragile object. 

\bigskip 
A brutal method was suggested in [AH92].  It consists of using explosive nuclear
radiations causing the sublimation of matter and therefore transverse impulses.  Our
experience at Aerospatiale shows that the explosion of a few megaton burst at a
distance of few kilometers causes a velocity increment given by the equation 

\smallskip 
\centerline {$ \displaystyle 
\Delta v \,=\, {{3 \, \pi}\over{8}} \, {{  I_0}\over{\rho \, R}} $}

\smallskip \noindent
where $\, \rho = 0.1  \,  {\rm ton/m}^3 \,$ is the typical density of a
comet, $\, R \,$  is its radius of 4 kms and $\, I_0 \,$   is the impulse induced by the
physical processes described above.  The typical value of $\, I_0 \,$  is around  3000
pascals.second, giving $\, \Delta v  = 9 \,\, 10^{-3} \,$  m/s, {\it i.e.} what
is desired.  It should be noted that the distance of some kilometers considered here
is quite unrealistic and other effects such as material shock waves inside the comet
may destroy the internal structure.  Moreover, a large amount of energy is lost in
space in the other directions by $\gamma$-ray emissions and we have not considered here
the ethical problem of using nuclear charges in the space environment. 

\bigskip 	
We conclude this section with a straightforward idea concerning impulse conservation. 
Let us assume that it is possible to sublimate a quantity of ice from the comet's
nucleus.  This water vapor would then expand in a vacuum with a rarefaction wave.  The
ejection velocity from the comet's nucleus is exactly equal to the speed of sound in
the gas (see {\it e.g.} the classic book by R.~Courant and K. Friedrichs [CF48]). 
Under these rarefied conditions of low external pressure and temperature, it is
difficult to estimate the speed of sound;  we suggest a speed $\, c = 100 $  meters per
second.  Then, the momentum conservation equation $\, M_2 \, \Delta v = m \, c \,$ 
establishes that the sublimation of $\, m =  3 \, \, 10^6 \,$  tons of Halley's comet will
deflect it by  $\, \Delta v = 1 \,$  cm/s.  The advantage of this approach is that no
propellant is required.  The drawback is that the ejection velocity is relatively low
and the energy cost of such a propulsion system is very high.

\bigskip 
From conventional thermodynamic data ({\it e.g.} [PR93]), the latent energy of fusion is
equal to 330 kjoules/kg and the evaporation energy is around 2300 kjoules/kg.  As a
first approach, we can estimate that the energy required for water sublimation is
equal to around $3 \, 10^6 $   joules/kilogram.  This means that the energy required
to create the appropriate impulse is $\, S_2 = 10^{16} $  joules, to be compared
with $\, E_2 = 10^9 $   joules and  $\, W_2 = 4 \,\, 10^{22} $  joules.  If
such an energy is communicated by a laser source, we could also take into account the
radiation pressure associated with the optical reflection of the ice off the comet,
creating an additional impulse. 

\bigskip 	
In this section, we have shown that human action is possible on a small asteroid using
today's technology one year before impact, with the help of gravity, which would
typically amplify the effect by a factor of 20.  The deposit of a mass of 10 metric
tons of propellant is sufficient to deflect a 100 m stony asteroid.  With the same
assumptions concerning kinetics, an energy of some megatons on a Halley-type comet
would probably create a sufficient impulse by sublimation.  This system design could
be an alternative to the explosion of a large nuclear charge at a short distance from
the meteor and should be studied in greater detail.

\bigskip    \noindent {\bf \large   5)	\quad Conclusion}

\smallskip 
We have studied the problem of possible bolide interception from three aspects: space
threat, global effects and human action.  The space threat is due to predictable
asteroids which are not sufficiently well known and to irregular comets whose possible
long term impact prediction remains generically uncertain.  The effects on Earth must
be considered globally using, for instance, methodologies developed by the Global
Change science community, taking into account the fact that the energy transfer
induced in the system by the impact of a large object is of the same order of
magnitude as the total energy dissipated each year by geophysical phenomena.  Finally,
possible human action remains modest in terms of velocity transfers (1 cm/s) when
using early detection, interception a long time before the predicted date of the
catastrophe and deflection with the help of the gravitational field.  Action using a
conventional propulsion system is possible for deflecting small asteroids, whereas a
very large source of energy would be necessary to change the trajectory of a
Halley-type comet by sublimation of a large quantity of solid ice of the comet's
nucleus. 

 \bigskip     \noindent {\bf  \large  Acknowledgments}

\smallskip 
The author thanks Jean Jouzel of the Commissariat \`a l'Energie
Atomique (Saclay) for providing reference [Koe96],  Max Calabro, Andr\'e Cariou,
Robert Clerc, Philippe Couillard, Jean Dupont and Jean-Bernard Renard of Aerospatiale
Espace \& Defense (Les Mureaux) for encouragements, stimulating discussions and active
participation on the subject presented herein.

 \bigskip     \noindent {\bf  \large  References }

\smallskip \hangindent=7mm \hangafter=1 \noindent
[AH92] Ahrens, A. Harris,
`` Deflection and Fragmentation of Near-Earth Asteroids'', {\it Nature}, volume 360, page  429–433,
December 3, 1992.

\smallskip \hangindent=7mm \hangafter=1 \noindent
[Alv80] L. Alvarez, W. Alvarez, F. Asaro, H. Michel, 
``Extraterrestrial Cause for the Cretace\-ous-Tertiary Extinction'', {\it Science}, volume 208,
pages 1095-1108, 1980. 

\smallskip \hangindent=7mm \hangafter=1 \noindent
[Aug92] J.F. Augereau,
``Le retour de Toutatis'', {\it Le Monde}, Paris, December 9, 1992. 

\smallskip \hangindent=7mm \hangafter=1 \noindent 
[BCR97] N. Bhandari, V. Courtillot, R. Rocchia,  
{\it European Congress of Geosciences}, March 1997.

\smallskip \hangindent=7mm \hangafter=1 \noindent
[CM94] C. Chapman, D. Morison,
``Impacts on  the Earth by Asteroids and Comets : Assessing the hazard'',
{\it Nature}, volume 367, pages 33-40,
January 6, 1994.

\smallskip \hangindent=7mm \hangafter=1 \noindent
[CF48] R. Courant, K. Friedrichs.
{\it Supersonic Flow and Shock Waves}, Interscience, New York, 1948.

\smallskip \hangindent=7mm \hangafter=1 \noindent
[CTZ93] C. Chyba, P. Thomas, K. Zahnle,
``The 1908 Tunguska Explosion: Atmospheric Disruption of a Stony Asteroid'', {\it Nature},
volume 361, pages 40-44, January 7, 1993.

\smallskip \hangindent=7mm \hangafter=1 \noindent
[Dar93] P. Darquey, 
``Hermes space plane. Design assessment report for meteorites and orbital debris environment'',
{\it Aerospatiale Espace} \& {\it Defense} n$^{\rm o}$ 120183, January 1993.

\smallskip \hangindent=7mm \hangafter=1 \noindent
[DF94] F. Duret, J.P. Frouard.
``Conception g\'en\'erale des syst\`emes spatiaux ; conception des fus\'ees porteuses'',
{\it Ecole Nationale Sup\'erieure de l'A\'eronautique et de l'Espace}, Toulouse, 1994.

\smallskip \hangindent=7mm \hangafter=1 \noindent
[Duf94] J.P. Dufour,
``La fin d'une \'etoile'',
{\it Le Monde}, Paris, July 16, 1994.

\smallskip \hangindent=7mm \hangafter=1 \noindent 
[Fra96] C. Frankel,  {\it La mort des dinosaures :
l'hypoth\`ese cosmique. Chronique d'une d\'ecouverte scientifique}, Masson, Paris,
April 1996.

\smallskip \hangindent=7mm \hangafter=1 \noindent
[GC96] W. Steffen, A. Shvidenko
(Eds). International Geosphere-Biosphere Programme : A Study of Global Change of the
International Council of Scientific Unions, {\it Report} n$^o$ 37, Stockholm, 1996.

\smallskip \hangindent=7mm \hangafter=1 \noindent
[GU61] R. Goscinny, A. Uderzo. 
{\it Asterix le Gaulois}, Dargaud, Paris, 1961.

\smallskip \hangindent=7mm \hangafter=1 \noindent
[HBC91] A. Hildebrand, W. Boynton, A. Camargo, G. Penfield, M. Pilkinson, D. Kring, 
``Chicxulub crater : a possible cretaneous-tertiary boundary impact crater on the
Yucatan peninsula, Mexico'', {\it Geology}, volume 19, pages 867-871, 1991.

\smallskip \hangindent=7mm \hangafter=1 \noindent
[Hir97] P. Le Hir, 
``Le choc probable d'une com\`ete et de Jupiter ``observ\'e'' 300 ans apr\`es'',
{\it  Le Monde}, Paris, January 16, 1997.

\smallskip \hangindent=7mm \hangafter=1 \noindent
[HS74] M. Hirsch, S. Smale,  {\it Differential Equations,
  Dynamical Systems, and Linear Algebra}, Academic Press, New York , 1974.

\smallskip \hangindent=7mm \hangafter=1 \noindent
[Koe96] C. Koeberl {\it et al.}. 
``Response of the Earth System to Impact Processes'', {\it Proposal for an European Science
  Foundation scientific programme}, November 1996.

\smallskip \hangindent=7mm \hangafter=1 \noindent
[Kul27] L. Kulik, 
``To the question about a place of the Tunguska meteorite fall'', 
{\it Doklady Akademii Nauk SSSR},  volume 23, pages 399-402, 1927. 

\smallskip \hangindent=7mm \hangafter=1 \noindent
[Mau89] J.L. Heudier, R. Chemin, A. Maury, C. Pollas, 
``1989AC'', {\it International Astronomical Union Circular} n$^{\rm o}$ 4701, page 1, January 5, 1989.

\smallskip \hangindent=7mm \hangafter=1 \noindent
[Oor63] J. Oort, 
``Empirical data on the origin of comets'',
in {\it The Moon, meteorites, and comets}, volume 4, {\it the solar system}, edited by
B.M. Middlehurst and G.P. Kuiper, U. of Chicago press, Chicago \& London, pages 665-673, 1963.

\smallskip \hangindent=7mm \hangafter=1 \noindent
[PR93] J. P\'erez, A. Romulus,  {\it Thermodynamique ;
fondements et applications},  Masson, Paris, 1993. 

\smallskip \hangindent=7mm \hangafter=1 \noindent
[Sag80] C. Sagan,  {\it Cosmos},  Random House, Inc, New York, 1980. 

\smallskip \hangindent=7mm \hangafter=1 \noindent
[SH96] P. Schultz, S. D'Hondt, 
``Cretaceous-Tertiary (Chicxulub) impact angle and its consequences'', 
 {\it Geology}, volume 24, pages 963–967,  November 1996.

\smallskip \hangindent=7mm \hangafter=1 \noindent
[Sho94] E. Shoemaker, C. Shoemaker,
``The crash of Shoemaher-Levy 9 into Jupiter and its implications for comet bombardment on Earth'', in
{\it New Developments Regarding the KT Event and Other Catastrophes in Earth History}, 
Houston University, pages 113-114, 1994. 

\smallskip \hangindent=7mm \hangafter=1 \noindent
[Spa94] Spacecast 2000, 
``Preparing for Planetary Defense.
Detection and Interception of Asteroids on Collision Course with Earth'', November 1994.

\smallskip \hangindent=7mm \hangafter=1 \noindent
[SRM91] J. Scotti, D. Rabinowitz, B. Marsden,
``Near miss of the Earth by a small asteroid'',            
{\it Nature}, volume 354, p. 287-289, 1991.

\smallskip \hangindent=7mm \hangafter=1 \noindent
[SSL93] E. Shoemaker, C. Shoemaker, D. Levy, J. Scotti,
P. Bendjoya, J. Mueller,  ``Comet Shoemaker-Levy (1993e)'', 
  {\it International Astronomical Union Circular} n$^{\rm o}$ 5725, page 1, March 26, 1993.

\smallskip \hangindent=7mm \hangafter=1 \noindent
[SWS90] E. Shoemaker, R. Wolf, C. Shoemaker, 
``Asteroids and comet flux in the neighborhood of the Earth'', 
{\it Geological  Society of America}, Special paper n$^{\rm o}$ 247, pages 155-170, 1990. 

\smallskip \hangindent=7mm \hangafter=1 \noindent 
[Tak95] T. Takata, 
``Three-dimensional analysis of impact processes on planets'',
PhD. Thesis, {\it  California Institute of Technology,}
NIPS-96-33178, Pasadena, 1995.

\smallskip \hangindent=7mm \hangafter=1 \noindent
[Whi30] F. Whipple,
``The great Siberian meteor and the waves, seismic and aerial, which it produced'' 
{\it  Quarterly Journal of the Royal Meteorological Society}, volume 56, pages~287-304, 1930. 

\smallskip \hangindent=7mm \hangafter=1 \noindent
[Whi63] F. Whipple, 
``On the origin of the cometary nucleous'',
in {\it The Moon, meteorites, and comets}, volume 4, {\it the solar system}, ed. by
B.M. Middlehurst and G.P. Kuiper, University of Chicago Press, Chicago \& London, p. 639-664, 1963.

\smallskip \hangindent=7mm \hangafter=1 \noindent 
[Wyc87] S. Wyckoff, 
``Ground-based obser-vations of Halley's comet'', {\it  AIAA paper} n$^{\rm o}$ 87-630, January 1987.

\fancyfoot[C]{\oldstylenums{\thepage}}

\end{document}